\documentstyle[12pt]{article}

\newcommand{\beq}{\begin{equation}}
\newcommand{\eeq}{\end{equation}}
\newcommand{\beqa}{\begin{eqnarray}}
\newcommand{\eeqa}{\end{eqnarray}}
\newcommand{\eq}[1]{(\ref{#1})}
\newcommand{\nn}{\nonumber}

\newcommand{\ra}{\rightarrow}
\newcommand{\trone}{{\rm tr}[1]}

\newcommand{\teta}{\Theta_2\Bigl(s{2\mu\over\beta},
is{4\pi\over\beta^2}\Bigr)}
\newcommand{\integ}{\int_0^\infty}

\newcommand{\NP}[1]{ {\it Nucl.~Phys.} {\bf #1}}

\newcommand{\PR}[1]{ {\it Phys.~Rev.} {\bf #1}}
\newcommand{\PRL}[1]{ {\it Phys.~Rev.~Lett.} {\bf #1}}

\newcommand{\IJMP}[1]{ {\it Int.~J.~Mod.~Phys.} {\bf #1}}

\newcommand{\ZP}[1]{ {\it Z.~Phys.} {\bf #1}}
\newcommand{\RMP}[1]{ {\it Rev.~Mod.~Phys.} {\bf #1}}
\begin{document}

\begin{titlepage}

\begin{flushright}
NBI-HE-97-30\\
hep-th/9707085
\end{flushright}

\setcounter{page}{0}
\vspace{5 mm}
\begin{center}
{\Large Note on thermodynamic fermion loop under \\
constant magnetic field}
\end{center} 
\vspace{10 mm}
\begin{center}
{\bf Haru-Tada Sato
\footnote{e-mail: sato@nbi.dk}} \\
\vspace{5mm}
{\small \em The Niels Bohr Institute, University of Copenhagen, \\
      Blegdamsvej 17, DK-2100 Copenhagen, Denmark}\\
\end{center}
\vspace{10mm}
\begin{abstract}
The one-loop effective potential of a thermodynamic fermion loop 
under constant magnetic field is studied. As expected, it can be 
interpreted literally as a discretized sum of $(D-2)$-dimensional 
energy density above the Dirac sea. Large/small mass expansions of 
the potential are also examined. 
\end{abstract}

\vfill

\begin{flushleft}
PACS: 11.30.Rd, 12.20.Ds \\
Keywords: magnetic field, dimensional reduction, effective potential
\end{flushleft}
\end{titlepage}
%%%%%%%%%%%%%%%%%%%%%%%%%%%%%%%%%%%%%%%%%%%%%%%%%%%%%%%%%%%%%%%%%%%%%
\renewcommand{\thefootnote}{\arabic{footnote}}
\setcounter{equation}{0}
\indent

Recently, effects of magnetic field on the chiral symmetry breaking 
have been intensively studied \cite{GMST}. In the paper \cite{KST}, 
we obtained a simple analytic form of thermodynamic effective 
potential for a fermion loop coupled to constant magnetic field in 
$D$ dimensions. On the contrary, corresponding ($D$-dimensional) 
effective potential for the electric field case \cite{Kli}-\cite{Baba} 
can not easily be organized into a concise form like the magnetic case 
due to a number of singularities; in other words, possessing an 
imaginary part \cite{imagin}. Nevertheless we derived a few analytic 
expressions for the real parts of the effective potential and its gap 
equation in the simplest (zero temperature) case through the 
electromagnetic dual transformation; i.e., rotating a magnetic 
parameter $\xi=eB$ into a pure imaginary number 
$\xi=i\rho=ieE$~\cite{KST}. This rotation procedure can not always 
be applied to an intermediate process of calculation, depending on 
whether or not the potential can be regarded as a complex analytic 
function of $\xi$. It is expected to be valid when the potential is 
written in terms of a special analytic function. In this way, a 
compact form of $D$-dimensional thermodynamic effective potential 
for the electric case might be obtained. 

Associated with this strategy, we here present some more analytic 
representations (in terms of special functions) of the 
thermodynamic effective potential for a fermion loop coupled to 
a constant magnetic field. First, we discuss a physical 
interpretation of the effective potential. As is well known, 
effective potentials possess the meaning of energy density, and 
we would like to suggest that a mathematical expression 
of the (finite temperature) potential indicates the physical 
meaning explicitly. There is also a dimensional reduction from 
$D$ to $D-2$ in the presence of magnetic field, and we hence expect 
that the structure should be $D-2$ dimensional energy density 
averaged by the Fermi-Dirac state density. 

We then present large/small mass expansions of the thermodynamic 
effective potential with keeping $\beta$ finite. In order to see a 
validity for these results, we shall reproduce a zero temperature 
potential (with $\mu\not=0$) from these expressions. Namely applying 
a low-temperature expansion to each mass expansion, we shall see 
that the leading terms of the low-temperature expansions coincide 
with the zero-temperature potential obtained in \cite{KST}, where 
none of these (mass and temperature) expansions were used and its 
$T=0$ situation was rather direct. In this sense, this paper provides 
another proof of the zero-temperature potential of finite density. 

The basic parts of model Lagrangian are 
\beq
{\cal L} ={\bar\psi}i\gamma^{\mu}(\partial_\mu - ieA_{\mu})\psi 
          -m {\bar\psi}\psi, 
\label{model}
\eeq
and the $D$-dimensional thermodynamic effective potential is derived 
at one loop \cite{KST} 
\beq
 V_{\beta,\mu}(m;\xi) =
    -  {\trone\over2\beta}
       \integ {ds\over s}{\teta\over(4\pi s)^{(D-1)/2}}
     s\xi {\rm coth}(s \xi) e^{-s(m^2-\mu^2)}\ , \label{Vbm} 
\eeq 
where $\Theta_2$ is the elliptic theta function of second kind and 
$\trone$ stands for the trace of gamma matrix unit. 
This can also be rewritten in the form 
\beq
 V_{\beta,\mu}(m;\xi) = 
V_R(m;\xi) + {\tilde V}_{\beta,\mu}(m;\xi) \ ,
\eeq
where the first term $V_R$ is a renormalized effective potential 
for the original Lagrangian \eq{model}, or one may add a four-fermi 
interaction to the Lagrangian \eq{model} if necessary. However, 
such a fermion interaction does not contribute to the second term 
${\tilde V}_{\beta,\mu}$, which is the pure thermodynamic part of 
the effective potential \eq{Vbm} and is renormalization free. 
Since our present interest is the thermodynamic part, which 
possesses a model independent structure, we focus on the latter 
potential, and its explicit form is given by \cite{KST} 
\beq
{\tilde V}_{\beta,\mu}(m;\xi)={\trone\xi\over(4\pi)^{D/2}}
\,\Bigl[\,{1\over2}{\cal O}_\beta(m)  
+ \sum_{l=1}^\infty{\cal O}_\beta(\sqrt{m^2+2l\xi}\,)\,\Bigr]\ ,
\label{discpote}
\eeq
where 
\beq
{\cal O}_\beta(\sigma)
= 4\sum_{n=1}^\infty(-1)^n\cosh(n\beta\mu)
\left({\beta n\over2\sigma}\right)^{1-D/2}K_{D/2-1}(n\beta\sigma)\ . 
\label{obetaK}
\eeq
Note that eq.\eq{discpote} is composed of the sum over a 
discrete energy spectrum $\epsilon_n=\pm\sqrt{m^2+2n\xi}$ due to 
the presence of magnetic field. 
Since $\xi$ possesses the mass dimension 2, the fundamental object 
${\cal O}_\beta(\sigma)$ in the pure thermal effective potential  
possesses mass dimension $D-2$. Hence eq.\eq{obetaK} can be 
interpreted as an effective potential in $D-2$ dimension. 
We can regard this fact as an example of the dimensional reduction 
from $D$ to $D-2$. Furthermore we here suggest one more interesting 
feature: From this specific potential form, we can read clearly the 
common fact that the effective potential is understood as an energy 
density. Although this statement is very natural, however the above 
potential forms eqs.\eq{Vbm} and \eq{obetaK} do not explain this 
structure explicitly. 

Let us reveal here the desired structure introducing another 
expression for eq.\eq{obetaK} 
\beq
{\cal O}_\beta(\sigma)={-2\sqrt{\pi}\over\Gamma({D-1\over2})}\,
\Bigl[\,\int_0^\infty {(t^2+2\sigma t)^{D-3\over2}\over
1+e^{\beta(t+\sigma+\mu)}}dt\quad +(\mu\ra-\mu)\,\Bigr]\ ,
\label{obeta}
\eeq
where we have used an integral representation for $K_{\nu}$ and 
then performed the sum on $n$. Changing the integration variable 
with $\epsilon=t+\sigma$, we obtain 
\beq
{\cal O}_\beta(\sigma)={-2\sqrt{\pi}\over\Gamma({D-1\over2})}\,
\Bigl[\,\int_\sigma^\infty {(\epsilon^2-\sigma^2)^{D-3\over2}\over
1+e^{\beta(\epsilon-\mu)}}d\epsilon\quad +(\mu\ra-\mu)\,\Bigr]\ .
\label{obeta2}
\eeq
RHS of this equation can be interpreted as follows. The denominator 
of the integrand is related to the state density $N(\epsilon)$ 
averaged by the Fermi-Dirac statistics 
\beq
N(\epsilon) = {1\over 1+e^{\beta(\epsilon-\mu)}} \ ,
\eeq
and the numerator with integration measure 
carrying the mass dimension $D-2$ (same as ${\cal O}_\beta$) 
\beq
f(\epsilon)d\epsilon = (\epsilon^2-\sigma^2)^{D-3\over2}d\epsilon \ ,
\eeq
is understood as an energy density for the infinitesimal interval 
$d\epsilon$. The integration means nothing but a total sum of this 
energy density over the whole state density $N(\epsilon)$, and the 
lower bound of the integration is restricted by the Dirac sea of the 
fermion mass $\sigma$. In addition, in the case of zero temperature, 
$N(\epsilon)$ takes 1 or 0, and the Fermi energy is given at the 
point $\epsilon=\mu$. Thus we have 
\beq
{\cal O}_\infty(\sigma)
\equiv \lim_{\beta\ra\infty}{\cal O}_{\beta}(\sigma) 
= \int_\sigma^\mu f(\epsilon)d\epsilon \ ,
\eeq
and this is exactly the result given in \cite{KST} 
\beq
{\cal O}_\infty(\sigma)
 =  -{2\sqrt{\pi}\over\Gamma({D+1\over2})}(2\sigma)^{D-3\over2}
(\mu-\sigma)^{D-1\over2} F\Bigl({3-D\over2},{D-1\over2};{D+1\over2};
{\sigma-\mu\over2\sigma}\Bigr)\,\theta(\mu-\sigma)\ . \label{oinf}
\eeq

Next, we consider further details of ${\cal O}_\beta(\sigma)$ 
(for finite $\beta$) which we did not get through in the previous 
paper \cite{KST}. We present below its large/small mass expansions 
at finite $\beta$. Obviously, in eq.\eq{obeta}, there are two ways 
of performing binomial expansion for the quantity $f(\epsilon)$; 
i.e., for $(t^2+2\sigma t)^\nu$ where $\nu=(D-3)/2$. 

First, let us consider the large $\sigma$ case 
($\sigma^{-1}$-expansion). After changing $\beta t \ra t$ and 
expanding around $\sigma=\infty$,we can perform the integration seen 
in eq.\eq{obeta} for each expansion mode
\beqa
{\cal O}_\beta(\sigma)&=&{-2\sqrt{\pi}\over\Gamma({D-1\over2})}\,
\beta^{2-D}\sum_{k=0}^\infty \left(\begin{array}{l}
 {D-3\over2} \\ \,\,\, k \end{array}\right)
(2\beta\sigma)^{{D-3\over2}-k}
\Gamma\Bigl({D-1\over2}+k\Bigr)  \nn \\
&\times&\exp\Bigl[-\beta(\sigma+\mu)\Bigr]\,
\Phi\Bigl({D-1\over2}+k;{1\over2} +
{i\beta\over2\pi}(\sigma+\mu),1\Bigr)+(\mu\ra-\mu)\ ,\label{imexp}
\eeqa
where $\Phi(z;s,a)$ is the Lerch transcendental function defined by 
\beq
\Phi(z;s,a)=\sum_{n=0}^\infty{e^{2\pi ins}\over(n+a)^z} 
= {1\over\Gamma(z)}\int_0^\infty
{e^{-at}t^{z-1}\over1-e^{-t+2\pi is}} dt 
\qquad (\mbox{Re}\,z>1,\quad 0<a\leq1). \label{Lerch}
\eeq
Since the Lerch function is transformed into the generalized 
zeta function by the relation 
\beq
e^{2\pi is}\Phi(1-z;s,1)=(2\pi)^{-z}\Gamma(z)\Bigl[
e^{{\pi i\over2}z}\zeta(z,s) +
e^{-{\pi i\over2}z}\zeta(z,1-s)\, \Bigr] \ , \label{LT}
\eeq
we can transform eq.\eq{imexp} into another representation
\beqa 
{\cal O}_\beta(\sigma)&=&{2\sqrt{\pi}\over\Gamma({D-1\over2})}\,
\sum_{k=0}^\infty \left(\begin{array}{l}
 {D-3\over2} \\ \,\,\, k \end{array}\right)
(2\sigma)^{{D-3\over2}-k}(2\pi)^{{D-3\over2}+k}
\Gamma\Bigl({D-1\over2}+k\Bigr) \Gamma\Bigl({3-D\over2}-k\Bigr)\nn\\
&\times&2\beta^{{1-D\over2}-k}\mbox{Re}\,\Bigl[
e^{{\pi i\over2}({3-D\over2}-k)}\,
\zeta\Bigl({3-D\over2}-k,{1\over2} +
{i\beta\over2\pi}(\sigma+\mu)\Bigr) + (\mu\ra-\mu)\Bigr] \ . 
\label{zetarep}
\eeqa
When $D=3$, the above summations in eqs.\eq{imexp} and \eq{zetarep}   
consist of only the $k=0$ term and 
the latter expression \eq{zetarep} contains superficial divergence 
(in the gamma function). 
In this sense, the former \eq{imexp} may rather be a fundamental 
expression. Anyway the both expressions coincide with the 
following in $D=3$ (see also the 1'st literature of \cite{GMST}) 
\beq
{\cal O}_\beta(\sigma) = -2{\sqrt{\pi}\over\beta} \ln
\Bigl[\,(1+e^{-\beta(\sigma+\mu)})
(1+e^{-\beta(\sigma-\mu)})\,\Bigr] \ .
\eeq
Nevertheless, the latter representation \eq{zetarep} is convenient 
to consider the low-temperature expansion ($\beta^{-1}$-expansion). 
Let us find that the leading term of the expansion reproduces the 
previous result \eq{oinf}. The $\beta^{-1}$-expansion can be 
performed through the asymptotic formula for zeta function
\beq
\zeta(z,a)\sim {1\over z-1}a^{1-z}\ , \qquad |a|\ra\infty\ .
\eeq
For example, the highest power of $\beta$ is calculated as follows:
\beqa
&&\mbox{Re}\Bigl[
e^{{\pi i\over2}({3-D\over2}-k)}\,
\zeta\Bigl({3-D\over2}-k,{1\over2} +
{i\beta\over2\pi}(\sigma\pm\mu)\Bigr)\Bigr] \nn\\
&&={1\over{1-D\over2}-k}
\Bigl({\beta\over2\pi}|\sigma\pm\mu|\Bigr)^{{D-1\over2}+k}
\mbox{Re}\,\Bigl[e^{{\pi i\over2}({3-D\over2}-k)}
e^{{\pi i\over2}(k+{D-1\over2})\mbox{sign}(\sigma\pm\mu)}\Bigr]
+\cdots \nn \\
&&={(\beta/2\pi)^{{D-1\over2}+k}\over{1-D\over2}-k}\times\left\{ 
\begin{array}{ll}
  0  &\quad\mbox{for} \quad +  \\
 \cos[\,(1-k-{D\over2})\pi\,] |\mu-\sigma|^{{D-1\over2}+k}
 \theta(\mu-\sigma)   &\quad \mbox{for} \quad - 
\end{array}\right. \label{expandzeta}\\
&&+\cdots\ . \nn
\eeqa
Substituting this result into RHS of eq.\eq{zetarep}, we derive 
\beq
{\cal O}_\beta(\sigma)={2\sqrt{\pi}\over\Gamma({D-1\over2})}\,
\sum_{k=0}^\infty \left(\begin{array}{l}
 {D-3\over2} \\ \,\,\, k \end{array}\right)
{(2\sigma)^{{D-3\over2}-k}\over{1-D\over2}-k}
(\mu-\sigma)^{{D-1\over2}+k}\theta(\mu-\sigma)
+{\cal O}(\beta^{-1})\ ,  \label{imbexp}
\eeq
where we have applied the following relation to eliminate the 
cosine factor seen in eq.\eq{expandzeta}
\beq
\Gamma({1\over2}+z)\Gamma({1\over2}-z)={\pi\over\cos\pi z}
\qquad \mbox{for}\quad z=1-k-{D\over2}.  \label{formula1}
\eeq
Furthermore, using
\beq
{1\over{1-D\over2}-k}={\Gamma({1-D\over2}-k)\over
\Gamma({3-D\over2}-k)}={(-1)^k\pi\over\cos(D{\pi\over2})
\Gamma({3-D\over2}-k)\Gamma({1+D\over2}+k)} \ ,
\eeq
and
\beq
F(\alpha,\beta;\gamma;z)=\sum_{n=0}^\infty{z^n\over n!}
{\Gamma(1-\alpha)\Gamma(1-\beta)\Gamma(\gamma)\over
\Gamma(1-\alpha-n)\Gamma(1-\beta-n)\Gamma(\gamma+n)}\ ,
\eeq
we find 
\beqa
{\cal O}_\beta(\sigma)&=&
{2\pi^{3\over2}(2\sigma)^{D-3\over2}(\mu-\sigma)^{D-1\over2}\over 
\cos(D{\pi\over2})\Gamma({3-D\over2})\Gamma({D-1\over2})
\Gamma({D+1\over2})}  \nn \\
&&\times F({3-D\over2},{D-1\over2};
{D+1\over2};{\sigma-\mu\over2\sigma})\theta(\mu-\sigma)
+{\cal O}(\beta^{-1})\ .
\eeqa
The leading term is now identified with the quantity \eq{oinf} 
through the formula \eq{formula1} for $z=(D-2)/2$. 

Secondly, let us compute the small mass expansion. The calculation 
is parallel to the above $\sigma^{-1}$-expansion. Expanding 
$(t^2+2\sigma t)^\nu$ for small $\sigma$ and integrating each mode, 
we obtain the following two representations 
\beqa
{\cal O}_\beta(\sigma)&=&{-2\sqrt{\pi}\over\Gamma({D-1\over2})}\,
\beta^{2-D}\sum_{k=0}^\infty \left(\begin{array}{l}
 {D-3\over2} \\ \,\,\, k \end{array}\right)
(2\beta\sigma)^k
\Gamma\Bigl(D-2-k\Bigr)  \nn \\
&\times&\exp\Bigl[-\beta(\sigma+\mu)\Bigr]\,
\Phi\Bigl(D-2-k;{1\over2} +
{i\beta\over2\pi}(\sigma+\mu),1\Bigr)+(\mu\ra-\mu)\ ,
\eeqa
and  
\beqa 
{\cal O}_\beta(\sigma)&=&{2\sqrt{\pi}\over\Gamma({D-1\over2})}\,
\sum_{k=0}^\infty \left(\begin{array}{l}
 {D-3\over2} \\ \,\,\, k \end{array}\right)
(2\sigma)^k (2\pi)^{{D-3-k}}
{\Gamma\Bigl({D-1-k\over2}\Bigr)\Gamma\Bigl({3-D+k\over2}\Bigr)
\over\cos[{\pi\over2}(3-D+k)]}\nn\\
&\times&\beta^{2-D+k}\mbox{Re}\,\Bigl[
e^{{\pi i\over2}(3-D+k)}\, \zeta\Bigl(3-D+k,{1\over2} +
{i\beta\over2\pi}(\sigma+\mu)\Bigr) + (\mu\ra-\mu)\Bigr] \ , 
\label{mexp}
\eeqa
where the Lerch transformation \eq{LT} and the following formula 
are applied
\beq
\Gamma(D-2-k)\Gamma(3-D+k)=
{\Gamma\Bigl({D-1-k\over2}\Bigr)\Gamma\Bigl({3-D+k\over2}\Bigr)
\over\cos[{\pi\over2}(3-D+k)]} \ .
\eeq
Again let us perform the $\beta^{-1}$-expansion in eq.\eq{mexp} 
and sum up the leading contribution similarly to the former case. 
The resulting equation is 
\beq
{\cal O}_\beta(\sigma)={-2\sqrt{\pi}\over\Gamma({D-1\over2})}\,
{(\mu-\sigma)^{D-2}\over D-2}
F({D-1\over2},D-1;3-D;{-2\sigma\over\mu-\sigma})
\theta(\mu-\sigma) +{\cal O}(\beta^{-1}) \ .
\label{mbexp}
\eeq
The value of ${\cal O}_\beta$ at the origin $\sigma=0$ for finite 
$\beta$ can be obtained by putting $\sigma=0$ into \eq{mexp}. 
The $k=0$ term is only nonzero, and we get ${\cal O}_\beta(0)$; 
\beq
{\cal O}_\beta(0) = {2\over\sqrt{\pi}}\left({2\pi\over\beta}\right)
^{D-2}\Gamma\left({3-D\over2}\right)\mbox{Re}\,\zeta\Bigl(3-D,
{1\over2}+i{\beta\mu\over2\pi}\Bigr)\ .         \label{obeta0}
\eeq
For infinite $\beta$, we just put $\sigma=0$ in \eq{mbexp}, and 
the leading term only survives; 
\beq
{\cal O}_\infty(0) = -{2\sqrt{\pi}\over\Gamma({D-1\over2})}
{\mu^{D-2}\over D-2} \ .  \label{zerovalue}
\eeq

In closing this report, we put two interesting remarks 
on eq.\eq{zerovalue}. The first one is that we can see that 
eq.\eq{zerovalue} is exactly a dimensionally reduced potential 
by $D\ra D-2$ up to a normalization constant factor. It can be 
seen by comparison with the potential of the continuum theory 
(i.e. $\xi=0$ case) \cite{IKM}: 
\beq
\lim_{\beta\ra\infty}{\tilde V}_{\beta,\mu}(0)=
 - 2 {\trone\over(4\pi)^{D/2}}
{2\sqrt{\pi}\over\Gamma({D+1\over2})} {\mu^D\over D} \ .
\eeq
Secondly, eq.\eq{zerovalue} is divergent if we take the limit 
$D\ra2$. This {\it is} an interesting point although a magnetic 
analysis in $D=2$ {\it per se} is meaningless. As pointed out 
in the introduction as well as in \cite{KST}, we can obtain an 
effective potential for the electric case at $T=0$ from the 
present case through the rotating $\xi\ra i\rho$. This rotation 
makes the quantity \eq{zerovalue} pure imaginary, and it certainly 
means that the imaginary part of the (electric case) potential is 
divergent in $D=2$ \cite{KKM}. 

%
%%%%%%%%%%%%%%%%%%%%%%%%%%%%%%%%%%%%%%%%%%%%%%%%%%%%%%%%%%%%%%%%%%%%%%%%
%                           REFERENCES                                 %
%%%%%%%%%%%%%%%%%%%%%%%%%%%%%%%%%%%%%%%%%%%%%%%%%%%%%%%%%%%%%%%%%%%%%%%%


\begin{thebibliography}{99}

\bibitem{GMST} V.P. Gusynin, V.A. Miransky and I.A. Shovkovy, 
       \PR{D52} (1995) 4718;\NP{B462} (1996) 249; \PRL{73} (1994) 3499;\\
        V.P. Gusynin and I.A. Shovkovy, hep-ph/9704394. 
\bibitem{KST} S. Kanemura, H-T. Sato and H. Tochimura, hep-ph/9707285.
\bibitem{Kli} K.G. Klimenko, Theor. Math. Phys. {\bf 89} (1992) 1287; 
               \ZP{C54} (1992) 323.
\bibitem{KKM} S. Kawati, A. Konisi and H. Miyata, \PR{D28} (1983) 1537.
\bibitem{KLL} S.P. Klevansky and R.H. Lemmer, \PR{D39} (1989) 3478.
\bibitem{NJLT}S.P. Klevansky, \RMP{64} (1992) 649.
\bibitem{Baba} A.Yu. Babansky, E.V. Gorbar and G.V. Shchepanyuk, 
             hep-th/9705218. 
\bibitem{imagin} M. Stone, \PR{D14} (1976) 3568;\\
                J. Schwinger, \PR{82} (1951) 664. 
\bibitem{IKM} T. Inagaki, T. Kouno and T. Muta, \IJMP{A10} (1995) 2241.
%
\end{thebibliography}
\end{document}